\documentclass[a4paper,11pt]{article}
\pdfoutput=1 % if your are submitting a pdflatex (i.e. if you have
\usepackage{jheppub} % for details on the use of the package, please
\usepackage[T1]{fontenc} % if needed

\usepackage{siunitx}	%for international units system
\usepackage{textcomp}	%for brakets \textlangle and \textrangle
\usepackage{graphicx}	%for \includegraphics{}

\title{\boldmath Spectroscopy of the tetraquark $c\bar{c}$-$c\bar{c}$ in a non-relativistic approach using a phenomenological QCD model}

\author[a,1]{J.A. Lesteiro-Tejeda,\note{Corresponding author.}}
\author[a]{D.A. Ramírez-Zaldívar}
\author[a]{C.E. García-Trápaga}
\author[a]{F. Guzmán-Martínez}
\affiliation[a]{Instituto Superior de Tecnologías y Ciencias Aplicadas (InSTEC), Universidad de La Habana,\\La Habana 10400, Cuba}
\emailAdd{jalesteiro@instec.cu}
\emailAdd{googol23nphy@gmail.com}
\emailAdd{cesar.garcia@instec.cu}
\emailAdd{guzman@instec.cu}
\abstract{Hadron spectroscopy is a powerful tool for testing the standard model and for the search of new physics. In this work, we create a tetraquark model from a di-meson interaction inspired by Jacobi’s coordinates. We consider mesons as thick points and quantify their interaction with the quark-antiquark interaction through a factor $\kappa$ ($\kappa\approx2$) and a central phenomenological potential, reducing the four-body problem finally to a one body equivalent problem. The eigenproblem is solved using a combination of the DVR method and perturbation theory in a C++ code. We obtain a mass spectrum for the tetraquark between \SI{6}{\giga\electronvolt} and \SI{8}{\giga\electronvolt} for the ground and first energy excitation level. The expected value $\sqrt{\textlangle r^2 \textrangle} < \SI{1}{\femto\meter}$ indicates that our system is compact. Finally, the critical case relation $v/c\approx0.3$ indicates that the non-relativistic approach used in all our formalism is valid.}

\begin{document} 
\maketitle
\flushbottom

\section{Introduction}
\label{sec:intro}
Tetraquark and pentaquark states were proposed as an explanation to exotics resonances detected in collider experiments. The discovery of X(3872) in 2003 by Belle and the posterior confirmation by Babar, CDFII, LHCb, and CMS \cite{Ali2017} established the state as genuine. The discovery pointed out a new direction in hadron spectroscopy, out of the B-meson decays were observed strongly interacting particles (four-quarks states), all containing a $c\bar{c}$ quark pair. These new proposed systems have been widely studied in two main currents: lattice-QCD calculations \cite{Bicudo2013,Braaten2013,Francis2017,Ikeda2014} and phenomenological QCD based effective potential models \cite{Anwar2018,Bai2016,Interaction1980,Brink1998,Debastiani2017,Debastiani2015,DeSouza2014,Ebert2009,Ebert2006,Ebert2008,Ebert2007,Ghalenovi2018,Gupta2012,Lloyd2004,Lucha1991}. Some examples of this exotic states are $X(3872)$, $Y(4140)$, $Y(4260)$, $Y(4360)$, $Y(4660)$, $Z_2(4250)$, $Z(4430)$ \cite{Ebert2010}, their masses are approximate to theoretical predictions of systems combining light and heavy quarks. The fact that full-heavy quark combinations have been studied to a lesser extent was our motivation to study spectroscopy and some properties of a $c\bar{c}-c\bar{c}$ tetraquark, along with the recent discovery of the state X(6900) \cite{RecienteJpsi} as a $J/\psi$-pair. We choose the effective potential method instead of lattice-QCD because it is less CPU power demanding and has proven useful to describe some particle's properties in hadron spectroscopy \cite{LichtenbergD.B.InstituteforNuclearTheoryandDepartmentofPhysics1987}.

\section{Quark-antiquark interaction and our tetraquark model}
\label{sec:interaction-and-model}
A potential will describe the quark-antiquark interaction (\ref{V_qq}) based on QCD phenomenology, which has a term called central (\ref{V_c}) and another called contact (\ref{V_s}). The central term considers the asymptotic freedom and the color confinement phenomenons, while the contact term is a gaussian smeared description of the quarks spin interactions. Similar potential models have been used to describe charmonium states successfully (\cite{Barnes2005,Debastiani2017}).

\begin{eqnarray}
\label{V_qq}
&V_{q\bar{q}}(r)=V_c+V_s&\\
\label{V_c}
&V_c = a/r + br&\\
\label{V_s}
&V_s = \frac{c}{m_qm_{\bar{q}}}(\frac{\sigma}{\sqrt{\pi}})^3e^{-\sigma^2r^2}\textbf{S}_{q} \cdot \textbf{S}_{\bar{q}}&
\end{eqnarray}

\begin{figure}[tbp]
	\centering
	\includegraphics[width=0.75\linewidth]{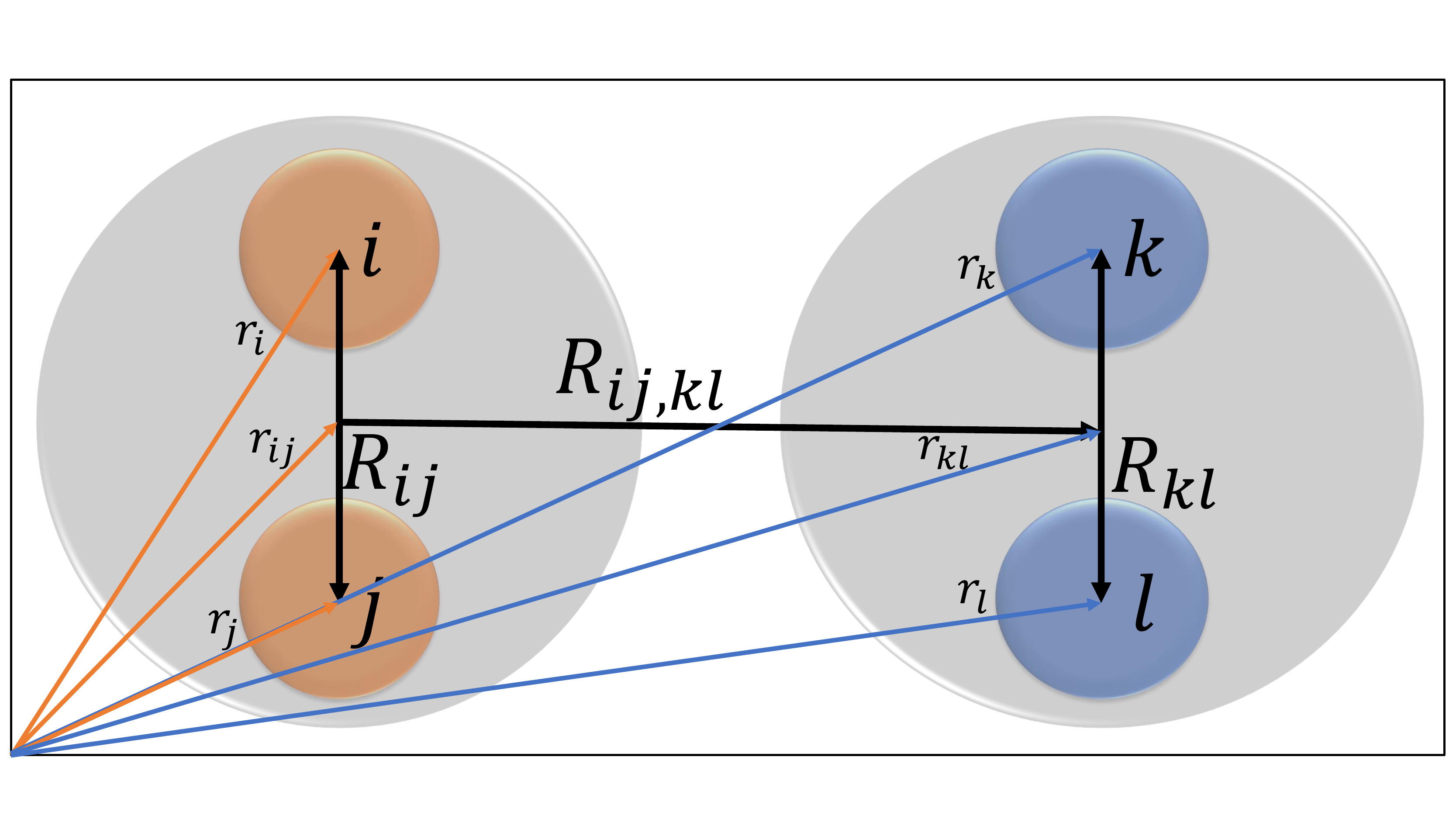}
	\caption{Tetraquark scheme.}
	\label{tq1}
\end{figure}

Our tetraquark model is inspired by the \textbf{H} Jacobi coordinates configuration (figure \ref{tq1}). The four-body Hamiltonian (\ref{H}) show under braced the terms corresponding to each meson and the effective inter-meson interaction. Given the central nature of the potential \ref{V_qq}, the one-body equivalent reduction is performed three times, one for each pair of quarks inside their respective meson and finally one more time for the mesons themselves (\ref{H_tres_un_cuerpo}). The term corresponding to the system’s center of mass is excluded from the Hamiltonian because the interaction with an external medium is not part of the present treatment.\\
To simplify the problem, we will assume the following. Instead of treating the quarks as punctual objects and the mesons resulting from quarks interaction, the mesons themselves will be punctual objects. The quark-antiquark interaction inside the meson will be present by replacing the reduced mass (\ref{mu}) with the reduced energy (\ref{Emu}) in expression \ref{H_tres_un_cuerpo}. The reduced energy will be computed with the energies eigenvalues from Schrödinger’s equation for the charmonium forming the tetraquark. So the mesons will be \emph{thick} points. Expression \ref{H_tetraquark} shows the result of applying $p=-i\hbar\nabla$ quantization rule.

\begin{equation}
\mathcal{H}=\underbrace{T_i+T_j+V_{ij}}_{q_i\bar{q}_j}+\underbrace{T_k+T_l+V_{kl}}_{q_k\bar{q}_l}
+\underbrace{V_{jl}+V_{jk}+V_{il}+V_{ik}}_{V_{ij,kl}}
\label{H}
\end{equation}

\begin{equation}
%\tiny
\mathcal{H}=\underbrace{\frac{1}{2}\mu_{ij}(\dot{\textbf{r}}_{ij})^2+V_{ij}(\textbf{r}_{ij})}_{q_i\bar{q}_j}%&\\ \nonumber
+\underbrace{\frac{1}{2}\mu_{kl}(\dot{\textbf{r}}_{kl})^2+V_{kl}(\textbf{r}_{kl})}_{q_k\bar{q}_l}%&\\ \nonumber
+\underbrace{\frac{1}{2}\mu_{ij,kl}(\dot{\textbf{R}}_{ij,kl})^2+V_{ij,kl}(\textbf{R}_{ij,kl})}_{q_i\bar{q}_jq_k\bar{q}_l}%&
\label{H_tres_un_cuerpo}
\end{equation}

\begin{equation}
\mu_{ij,kl}=\frac{(m_i+m_j)(m_k+m_l)}{(m_i+m_j)+(m_k+m_l)}
\label{mu}
\end{equation}

\begin{eqnarray}
\label{Emu}
&E_{\mu}=\frac{E_{ij}E_{kl}}{E_{ij}+E_{kl}}\rightarrow H_{ij}\Psi_{ij}=E_{ij}\Psi_{ij}&\\
&E_{ij}=m_i+m_j+\Delta E \nonumber&
\end{eqnarray}

\begin{equation}
\mathcal{H}\cong\frac{-\hbar^2\nabla^2_ {R_{ij,kl}}}{2E_{\mu}} + V_{ij,kl}(\textbf{R}_{ij,kl})
\label{H_tetraquark}
\end{equation}

\begin{figure}[tbp]
	\centering
	\includegraphics[width=0.75\linewidth]{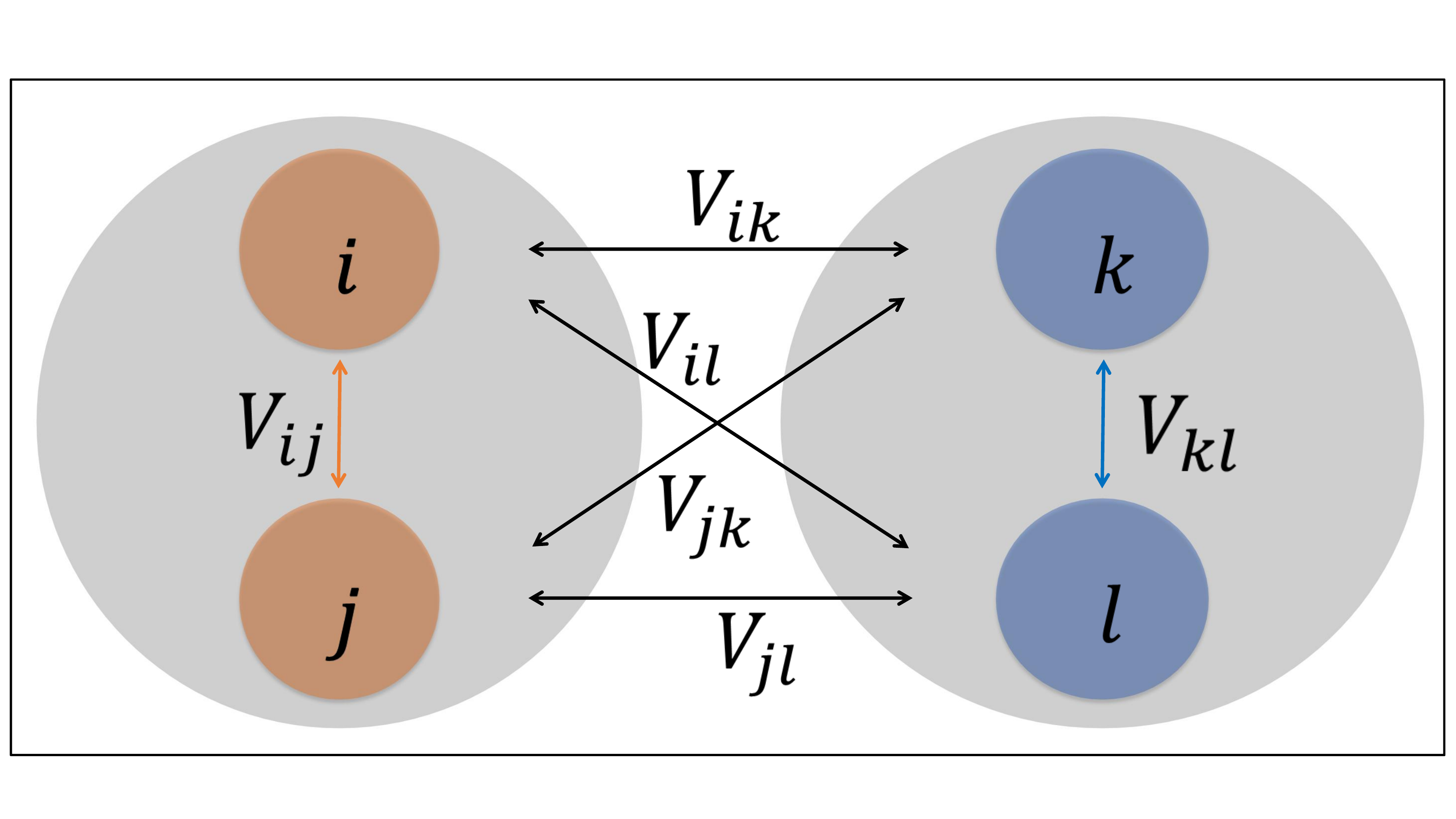}
	\caption{Independent quark interactions.}
	\label{V_k4}
\end{figure}

The two mesons inside the tetraquark can not exchange gluons because they are color-neutral in our mode. The inter-meson interaction has a residual nature, similar to protons and neutrons inside the atomic nucleus. We quantize this unknown interaction from the quark-antiquark interaction, which we can calculate. To achieve this, we will assume that quarks and antiquarks can interact individually (figure \ref{V_k4}), and not all of them need to be involved.

In this sense, expression \ref{H_tetraquark} is transformed into \ref{H_tetraquark2}, were $\kappa$ indicates how many quarks interact effectively, or in other words, how many quark-antiquark interactions worth the meson-meson interaction.

\begin{equation}
\mathcal{H}\cong\frac{-\hbar^2\nabla^2_ {R_{ij,kl}}}{2E_{\mu}} + \kappa V_{q\bar{q}}(R_{ij.kl})
\label{H_tetraquark2}
\end{equation}

\section{Wavefunctions, expected values and mass spectrum}
\label{sec:wavefunctions-mass}
\begin{figure}[tbp]
	\centering
	\includegraphics[width=0.75\linewidth]{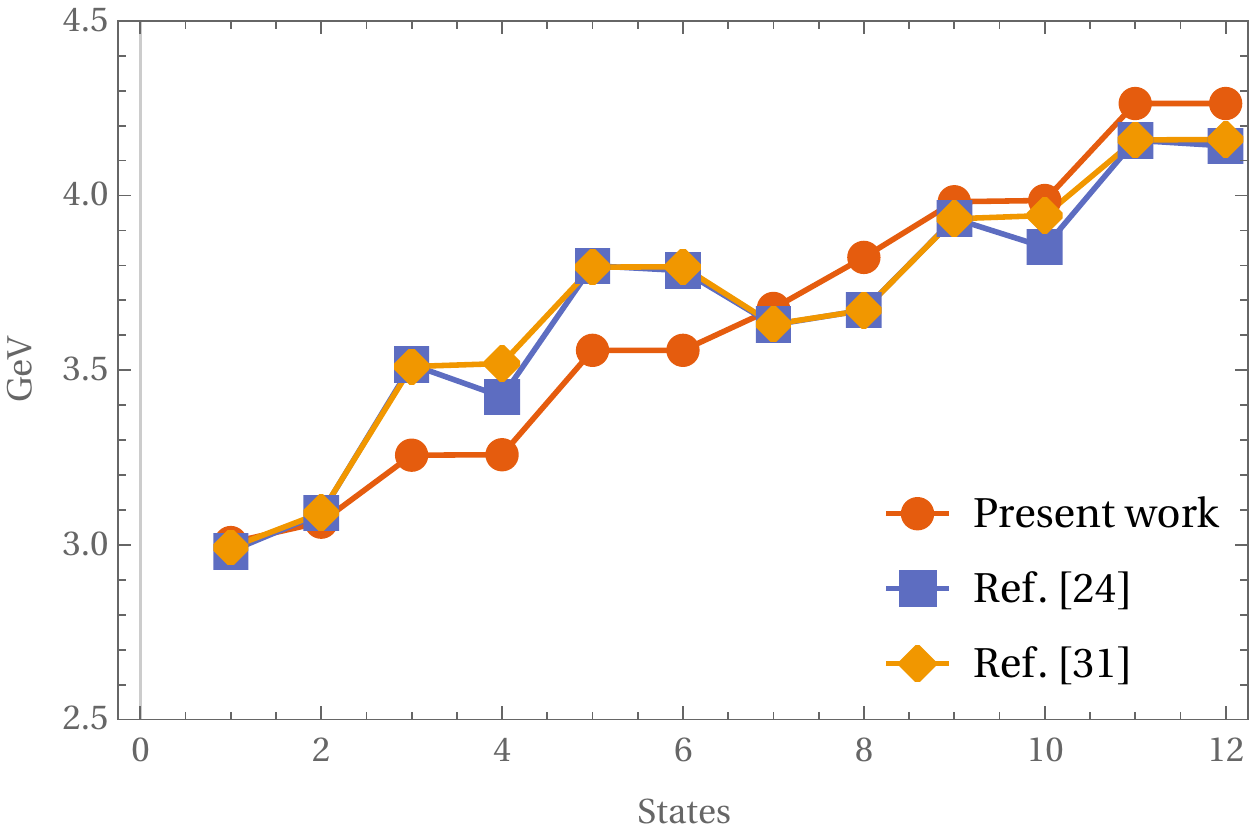}
	\caption{Mass spectrum of $c\bar{c} \ (\si{\giga\electronvolt})$.\label{espectro_c_barc_fig}}
\end{figure}

To solve the eigenproblem, we will use a combination of the DVR method \cite{Light1985,Colbert1991,D.A.RamirezZaldivar2018,Prudente2001} and perturbation theory \cite{Sakurai} in a C++ program written by the authors. Hamiltonian considering only the central part of the potential (\ref{V_c}) will be solved by DVR, which has proven to be a useful and relatively simple method for similar molecular potentials \cite{D.A.RamirezZaldivar2018,Colbert1991,Light1985,Prudente2001}. Later, the spin contribution to the potential (\ref{V_s}) will be considered a perturbation to the system, finally obtaining the perturbed eigenvalues and eigenfunctions.\\
The quark-antiquark interaction was parameterized from the charmonium experimental mass \cite{Family2018} and our energy levels are in good agreement also with previous calculations (table \ref{espectro_c_barc}, figure \ref{espectro_c_barc_fig}). The parameters are set as $a=\SI{3.2997}{}$, $b=\SI{0.5261}{\giga\electronvolt\squared}$, $c=\SI{4.1000}{}$ and $\sigma=\SI{15.9174}{\giga\electronvolt}$.

\begin{table}[tbp]
	\centering
	\begin{tabular}{|c|cccc|}
		\hline
		      State        & PDG \cite{Family2018} & Present work & Ref. \cite{Barnes2005} & Ref. \cite{Debastiani2019} \\
		        %          &                       &              &                        &                            \\
		$1\ ^1S\ (\eta_c)$ &        2.9839         &    3.0067    &         2.982          &           2.9924           \\
		$1\ ^3S\ (J/\psi)$ &        3.0969         &    3.0649    &         3.090          &           3.0917           \\
		 $1\ ^1P\ (h_c)$   &        3.5252         &    3.2567    &         3.516          &           3.5105           \\
		 $1\ ^3P\ (X_c)$   &        3.4147         &    3.2580    &         3.424          &           3.5191           \\
		     $1\ ^1D$      &           -           &    3.5569    &         3.799          &           3.7951           \\
		     $1\ ^3D$      &           -           &    3.5569    &         3.785          &           3.7958           \\
		$2\ ^1S\ (\eta_c)$ &        3.6376         &    3.6768    &         3.630          &           3.6317           \\
		 $2\ ^3S\ (\psi)$  &        3.6861         &    3.8229    &         3.672          &           3.6714           \\
		     $2\ ^1P$      &           -           &    3.9822    &         3.934          &           3.9334           \\
		     $2\ ^3P$      &           -           &    3.9858    &         3.852          &           3.9427           \\
		     $2\ ^1D$      &           -           &    4.2633    &         4.158          &           4.1591           \\
		     $2\ ^3D$      &           -           &    4.2633    &         4.142          &           4.1604           \\ \hline
	\end{tabular}
	\caption{Mass spectrum of $c\bar{c} \ (\si{\giga\electronvolt})$.}\label{espectro_c_barc} 
\end{table}

The kappa value is needed for calculating the tetraquark´s wavefunctions and mass spectrum and remains unknown. To find it, a fit will be made to both a theoretical prediction \cite{Debastiani2017} and a recently discovered state by LHCb collaboration $X(6900)$ \cite{RecienteJpsi}. We will use both values to fit because there are not enough experimental results or theoretical predictions for this system. Both cases show similar results: $\kappa=2.038$ when fitting from \cite{Debastiani2017}, and $\kappa=2.377$ fitting from \cite{RecienteJpsi}. Interestingly, $\kappa$ takes a fractional value between 2 and 3, different from the entire value we expect from the reasoning that leads us to figure \ref{V_k4}. The explanation for this within our model lies either on the strong interaction's anti-screening phenomena because the gluons can interact with themselves or that a binary approximation is not accurate enough for the four quark collective interaction problem.

We validate the non-relativistic approach for our tetraquark system from the value of the relation $v/c$, where $c$ is the velocity of light in the vacuum, and $v$ is the relative speed of composing quarks. Values close to 0 validate the non-relativistic approach while values close to 1 do not. Since kinetic energy operator is a quantum version of the non-relativistic expression \ref{Tclassica}, dimensional analysis and the equivalent of the Virial theorem for quantum mechanics \ref{virial} \cite{Sakurai}, lead us to \ref{vc}. Expression \ref{vcnumeric} is how we performed the numerical evaluation. 

\begin{equation}
T=\frac{p²}{2m}
\label{Tclassica}
\end{equation}
\begin{equation}
\textlangle T\textrangle=\textlangle \textbf{x}\cdot \nabla V\textrangle
\label{virial}
\end{equation}
\begin{equation}
v/c=\frac{\textlangle T\textrangle}{2m}
\label{vc}
\end{equation}
\begin{equation}
\textlangle \textbf{x} \cdot \nabla V \textrangle = 
\frac{\int r \cdot \frac{dV(r)}{dr} dr}
{\int r dr} = 
\frac{\int r \cdot dV(r)}{\int r dr}
\label{vcnumeric}
\end{equation}

\begin{table}[tbp]
	\centering
	\begin{tabular}{|c|cc|cc|cc|}
		\hline
		State  &        Mesons         & $v/c_0$ &        Mesons         & $v/c_0$ &        Mesons         & $v/c_0$ \\
		$1\ ^1S$   & $\ (1\ ^1S)\ (1\ ^1S)$ & 0.2982  & $\ (2\ ^1S)\ (2\ ^1S)$ & 0.2439  & $\ (3\ ^1S)\ (3\ ^1S)$ & 0.2094  \\
		$1\ ^3S$   & $\ (1\ ^1S)\ (1\ ^3S)$ & 0.2953  & $\ (2\ ^1S)\ (2\ ^3S)$ & 0.2392  & $\ (3\ ^1S)\ (3\ ^3S)$ & 0.2037  \\
		$1\ ^5S$   & $\ (1\ ^3S)\ (1\ ^3S)$ & 0.2924  & $\ (2\ ^3S)\ (2\ ^3S)$ & 0.2345  & $\ (3\ ^3S)\ (3\ ^3S)$ & 0.1979  \\
		$1\ ^1P$   & $\ (1\ ^1S)\ (1\ ^1P)$ & 0.2633  & $\ (2\ ^1S)\ (2\ ^1P)$ & 0.2188  & $\ (3\ ^1S)\ (3\ ^1P)$ & 0.1895  \\
		$1\ ^3P$   & $\ (1\ ^3S)\ (1\ ^1P)$ & 0.2609  & $\ (2\ ^3S)\ (2\ ^1P)$ & 0.2147  & $\ (3\ ^3S)\ (3\ ^1P)$ & 0.1844  \\
		$1\ ^5P$   & $\ (1\ ^3S)\ (1\ ^3P)$ & 0.2608  & $\ (2\ ^3S)\ (2\ ^3P)$ & 0.2146  & $\ (3\ ^3S)\ (3\ ^3P)$ & 0.1843  \\ \hline
	\end{tabular}
	\caption{Relation $v/c$ for tetraquark states with different compositions.\label{virialtabla}}
\end{table}

Table \ref{virialtabla} confirms this relation is inversely proportional to the state's mass. Therefore, as the critical case, 
\begin{displaymath}
v/c=0.2982,
\end{displaymath}
we conclude that the non-relativistic approach is valid while useful for satisfactory results studying our system.

\begin{figure}[tbp]
	\centering
	\includegraphics[width=0.75\linewidth]{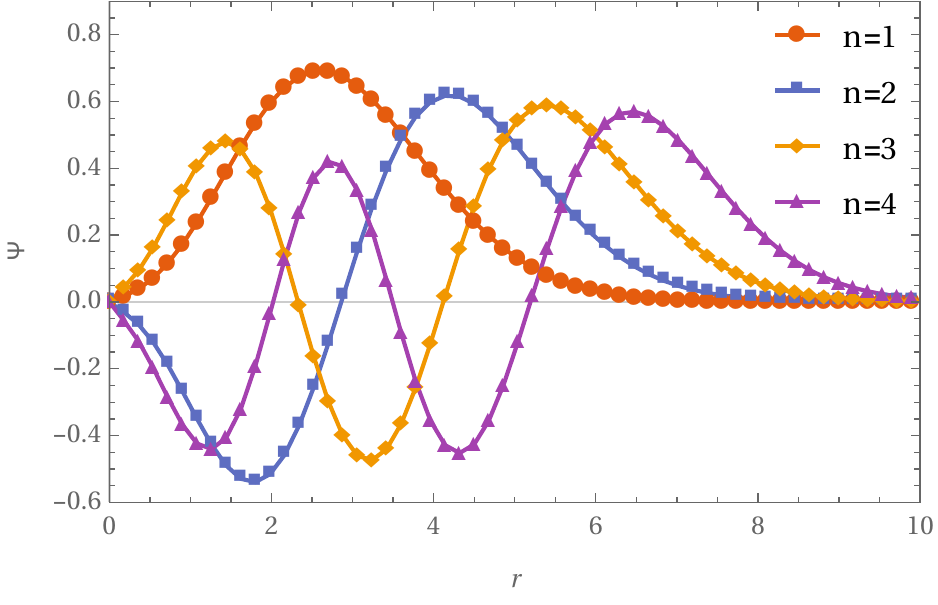}
	\caption{ Di-$J/\psi$ wave function.}\label{tq_wf_JPsi}
\end{figure}

The wave functions shown in \ref{tq_wf_JPsi} correspond to a di-$J/\psi$ tetraquark system, $n^5S$ where $n\epsilon[1,4]$ and $\kappa=2.038$. These are continuous wave functions like their first derivatives, and the number of zeros is one unit less than the energetic excitation level. For other tetraquark states and $\kappa=2.377$, the results are similar. Table \ref{valores_esperados} contains the expected values of some magnitudes of interest for the system.

\begin{table}[tbp]
	\centering
	\begin{tabular}{|c|ccc|ccc|}
		\hline
		State	&\multicolumn{3}{c|}{$\kappa=\SI{2.038}{}$}&\multicolumn{3}{c|}{$\kappa=\SI{2.377}{}$}\\
		$c\bar{c}-c\bar{c}$&$\textlangle r \textrangle$&$\textlangle 1/r \textrangle$&$\sqrt{\textlangle r^2 \textrangle}$&$\textlangle r \textrangle$&$\textlangle 1/r \textrangle$&$\sqrt{\textlangle r^2 \textrangle}$\\
		& & & & & &  \\
		$1 ^1S$ &0.5404 &2.0671 &0.5650 &0.5342 &2.0761 &0.5572 \\
		$1 ^3S$ &0.5404 &2.0673 &0.5650 &0.5342 &2.0763 &0.5572 \\
		$1 ^5S$ &0.5404 &2.0673 &0.5650 &0.5342 &2.0762 &0.5572 \\
		$1 ^1P$ &0.5579 &1.9849 &0.5817 &0.5494 &2.0045 &0.5717 \\
		$1 ^3P$ &0.5577 &1.9857 &0.5815 &0.5492 &2.0052 &0.5716 \\
		$1 ^5P$ &0.5577 &1.9857 &0.5815 &0.5492 &2.0052 &0.5716 \\
		& & & & & &  \\
		$2 ^1S$ &0.6862 &1.9068 &0.7427 &0.6700 &1.9250 &0.7235 \\
		$2 ^3S$ &0.6861 &1.9076 &0.7426 &0.6699 &1.9256 &0.7234 \\
		$2 ^5S$ &0.6861 &1.9073 &0.7426 &0.6699 &1.9254 &0.7234 \\
		$2 ^1P$ &0.7033 &1.8236 &0.7584 &0.6849 &1.8522 &0.7373 \\
		$2 ^3P$ &0.7031 &1.8246 &0.7583 &0.6847 &1.8530 &0.7371 \\
		$2 ^5P$ &0.7031 &1.8245 &0.7583 &0.6847 &1.8530 &0.7371 \\
		\hline
	\end{tabular}
	\caption{Expected values \big(r (\si{\femto\meter})\big)\label{valores_esperados}.}
\end{table}

The magnitude $\sqrt{\textlangle r^2 \textrangle}$ is directly related to the width of the gaussian wave packet \cite{Sakurai}. Therefore, our model results in a compact tetraquark system, $\sqrt{\textlangle r^2 \textrangle} < \SI{1}{\femto\meter}$.

\begin{figure}[tbp]
	\centering
	\includegraphics[width=0.75\linewidth]{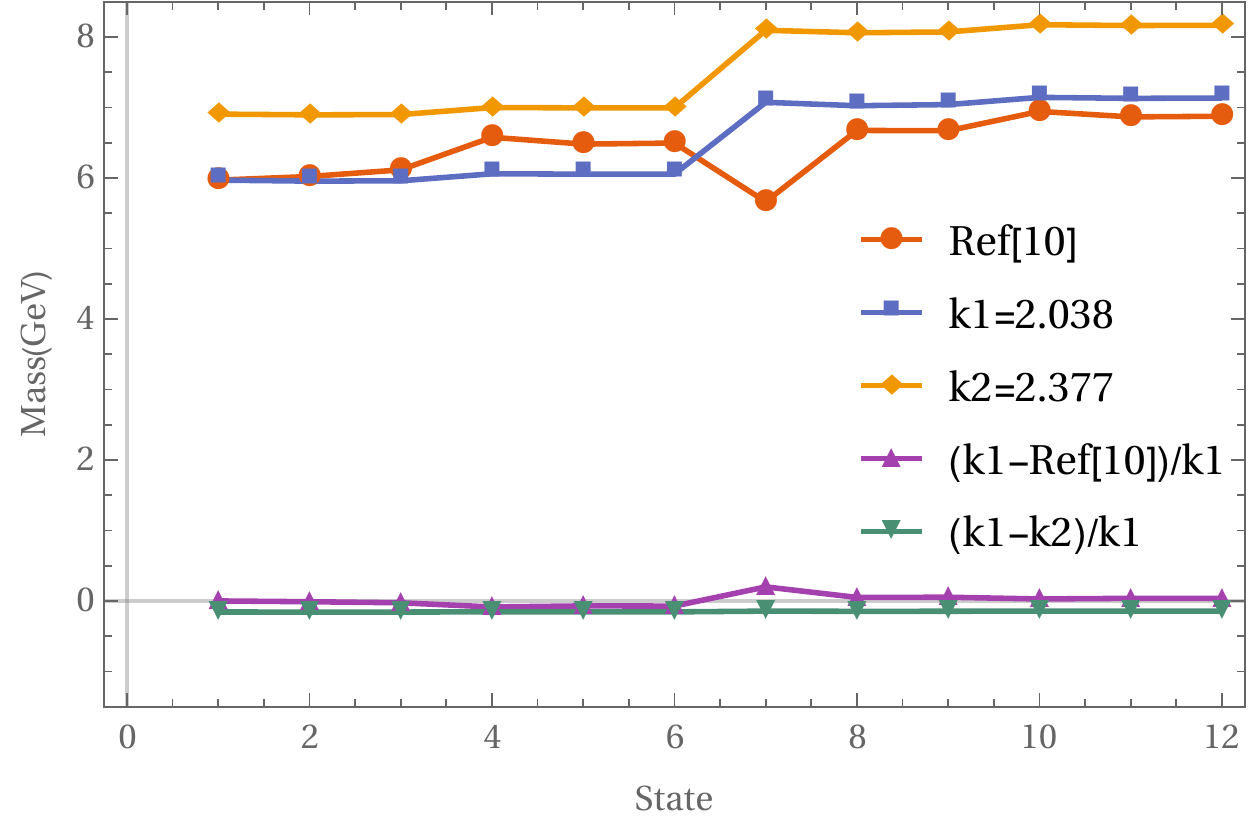}
	\caption{Mass spectrum of $c\bar{c}c\bar{c} \ (\si{\giga\electronvolt})$ and relative differences.\label{tqmiovsdebastianifig}}
\end{figure}

Table \ref{tqmiovsdebastianitabla} and figure \ref{tqmiovsdebastianifig} show the mass spectrum calculated for the tetraquark using both approaches already mentioned.  Our calculations show a tendency to increase the mass with an increase in energy excitation level $n$ and orbital momentum $l$, while decreasing mass when increasing spin momentum excitation $s$. Other theoretical predictions place the spectrum between \SI{6}{\giga\electronvolt} and \SI{9}{\giga\electronvolt} \cite{Lloyd2004} and \SI{6}{\giga\electronvolt} and \SI{7}{\giga\electronvolt} \cite{Chen2018}. These results are achievable with $\kappa\epsilon[2,3]$ in our model.

\begin{table}[tbp]
	\centering
	\begin{tabular}{|cc|cccc|}
		\hline
		         &                  & \multicolumn{4}{|c|}{Mass}                                                        \\
		 State   &   Composition    & Ref\cite{Debastiani2017} & k = 2.038 & Ref\cite{RecienteJpsi} & k = 2.377 \\
		$1\ ^1S$ & $\eta_c\ \eta_c$ &            5.9694            &   INPUT   &             -              &  6.9057   \\
		$1\ ^3S$ & $\eta_c\ J/\psi$ &            6.0209            &  5.9550   &             -              &  6.8956   \\
		$1\ ^5S$ & $J/\psi\ J/\psi$ &            6.1154            &  5.9598   &           6.9000           &   INPUT   \\
		$1\ ^1P$ &  $\eta_c\ h_c$   &            6.5771            &  6.0591   &             -              &  6.9986   \\
		$1\ ^3P$ &  $J/\psi\ h_c$   &            6.4804            &  6.0543   &             -              &  6.9948   \\
		$1\ ^5P$ &  $J/\psi\ X_c$   &            6.4954            &  6.0556   &             -              &  6.9957   \\
		$2\ ^1S$ & $\eta_c\ \eta_c$ &            5.6633            &  7.0731   &             -              &  8.0950   \\
		$2\ ^3S$ & $\eta_c\ J/\psi$ &            6.6745            &  7.0251   &             -              &  8.0597   \\
		$2\ ^5S$ & $J/\psi\ J/\psi$ &            6.6698            &  7.0411   &             -              &  8.0715   \\
		$2\ ^1P$ &  $\eta_c\ h_c$   &            6.9441            &  7.1434   &             -              &  8.1737   \\
		$2\ ^3P$ &  $J/\psi\ h_c$   &            6.8665            &  7.1295   &             -              &  8.1624   \\
		$2\ ^5P$ &  $J/\psi\ X_c$   &            6.8756            &  7.1339   &             -              &  8.1659   \\ \hline
	\end{tabular}
	\caption{Mass spectrum of $c\bar{c}c\bar{c} \ (\si{\giga\electronvolt})$.\label{tqmiovsdebastianitabla}}
\end{table}

\section{Concluding remarks}
\label{sec:conc}
The diquark interaction model we elaborated inspired by Jacobi coordinates, can consider angular excitation, both orbital and spin. The $\kappa$ value we introduced to quantify the meson-meson interaction is $\kappa=2.038$ fitting from theoretical predictions and $\kappa=2.377$ fitting from LHCb di-$J/\psi$ mass measurement. Both results are similar considering their implications: the tetraquark could be a quark bound to an exotic hadron or a configuration of binary interactions among the four quarks. Either way, the excess over ideally $\kappa=2$ reflects an anti-shadowing of the strong interaction or is a consequence of using a binary interaction approximation for the four-body problem. 
With our model, we obtain a mass spectrum similar to other theoretical predictions and experimental evidence, with masses between \SI{6}{\giga\electronvolt} and \SI{8}{\giga\electronvolt} approximately for the first and second excitation levels. The non-relativistic approach used in charmonium-like systems is also valid for our di-charmonium tetraquark: $v/c\approx0.3$ in the critical case. Furthermore, our tetraquark turns out to be a compact system since $\sqrt{\textlangle r^2 \textrangle} < \SI{1}{\femto\meter}$. The latter magnitude is related to the width of the gaussian wave packet.

\bibliography{bib/my.bib}

\end{document}